\documentclass[amsmath,amssymb,twocolumn,prl,floatfix,showpacs,nofootinbib]{revtex4-2}
\usepackage{amsmath}
\usepackage{graphicx}
\usepackage{float}
\usepackage{subfigure}
\usepackage{epstopdf}
\usepackage{color}
\usepackage{multirow}
\usepackage{setspace}
\usepackage{overpic}
\usepackage{amssymb}
\usepackage[colorlinks,urlcolor=blue, citecolor=blue,linkcolor=blue] {hyperref}
\usepackage{lineno}
\usepackage{bm}
\usepackage{rotating}
\usepackage{array}
\usepackage{booktabs}
\usepackage[utf8]{inputenc}
\let\oldequation\equation
\let\oldendequation\endequation
\renewenvironment{equation}
 {\linenomathNonumbers\oldequation}
 {\oldendequation\endlinenomath}

\begin{document}

\title{\bf \boldmath
Evidence for Two Excited $\Omega^{-}$ Hyperons
}

\author{
\begin{small}
\begin{center}
M.~Ablikim$^{1}$, M.~N.~Achasov$^{4,c}$, P.~Adlarson$^{76}$, O.~Afedulidis$^{3}$, X.~C.~Ai$^{81}$, R.~Aliberti$^{35}$, A.~Amoroso$^{75A,75C}$, Y.~Bai$^{57}$, O.~Bakina$^{36}$, I.~Balossino$^{29A}$
% ...
\\
\vspace{0.2cm}
(BESIII Collaboration)\\
\vspace{0.2cm} {\it
$^{1}$ Institute of High Energy Physics, Beijing 100049, People's Republic of China\\
% ...
$^{3}$ Bochum  Ruhr-University, D-44780 Bochum, Germany\\
$^{4}$ Budker Institute of Nuclear Physics SB RAS (BINP), Novosibirsk 630090, Russia\\
% ...
$^{29}$ INFN Sezione di Ferrara, (A)INFN Sezione di Ferrara, I-44122, Ferrara, Italy; (B)University of Ferrara,  I-44122, Ferrara, Italy\\
% ...
$^{35}$ Johannes Gutenberg University of Mainz, Johann-Joachim-Becher-Weg 45, D-55099 Mainz, Germany\\
$^{36}$ Joint Institute for Nuclear Research, 141980 Dubna, Moscow region, Russia\\
% ...
$^{57}$ Southeast University, Nanjing 211100, People's Republic of China\\
% ...
$^{75}$ University of Turin and INFN, (A)University of Turin, I-10125, Turin, Italy; (B)University of Eastern Piedmont, I-15121, Alessandria, Italy; (C)INFN, I-10125, Turin, Italy\\
$^{76}$ Uppsala University, Box 516, SE-75120 Uppsala, Sweden\\
% ...
$^{81}$ Zhengzhou University, Zhengzhou 450001, People's Republic of China\\
\vspace{0.2cm}
% ...
$^{c}$ Also at the Novosibirsk State University, Novosibirsk, 630090, Russia\\
% ...
}
\end{center}
\end{small}
}

\begin{abstract}
Using $e^+e^-$ collision data corresponding to an integrated luminosity of 19\,fb$^{-1}$ collected by the BESIII detector at center-of-mass energies ranging from 4.13 to 4.70\,GeV,
we report the first evidence for a new excited $\Omega^{-}$ hyperon, 
the $\Omega(2109)^{-}$, through the process $e^+ e^- \to \Omega(2109)^{-} \bar{\Omega}^{+} +c.c.$ 
with a significance of 4.1\,$\sigma$.
The mass and width of $\Omega(2109)^{-}$ are measured to be
$2108.5 \pm 5.2_{\rm stat} \pm 0.9_{\rm syst}\,{\rm MeV}/c^{2}$
and
$18.3 \pm 16.4_{\rm stat} \pm 5.7_{\rm syst}\,{\rm MeV}$, respectively.
We also present evidence for a new production mechanism for the previously identified $\Omega(2012)^-$
via the process $e^+ e^- \to \Omega(2012)^{-} \bar{\Omega}^{+} +c.c.$ with a significance of 3.5\,$\sigma$.
\end{abstract}

\maketitle

\oddsidemargin  -0.2cm
\evensidemargin -0.2cm


Baryons are the simplest system in which all three colors of QCD form a colorless object.  Understanding their structure and spectroscopy is a necessary component of a real understanding of QCD.
However, many fundamental issues in baryon spectroscopy are still not well
understood~\cite{Capstick:2000dk, baryonspectrum}.
One of the most important problems among them is the missing excited baryons: the number of observed baryon excited states is significantly smaller than the number predicted in the quark model~\cite{baryonspectrum, understandbaryon, diquark}. 
Thus, our present understanding of baryon spectroscopy is clearly incomplete.
Compared with other light baryon spectra, our knowledge of the $\Omega^{-}$ hyperon spectrum~\cite{pdg2024} is still very sparse, motivating further searches for $\Omega^{-}$ hyperon excited states. 
Before 2018, the Particle Data Group (PDG)~\cite{pdg2018} only listed three $\Omega^{-}$ resonances beyond the ground state $\Omega^{-}$, i.e., the $\Omega(2250)^-$, $\Omega(2380)^-$, and $\Omega(2470)^-$, and none of these three are considered well-established. 
The Skyrme model~\cite{Oh:2007cr}, the quark model~\cite{Capstick:1986ter, Faustov:2015eba, Chao:1980em, Chen:2009de, Pervin:2007wa, Liu:2019wdr}, 
the models with hyperfine interactions between quarks~\cite{An:2013zoa, An:2014lga}, and lattice QCD~\cite{Engel:2013ig, Edwards:2012fx, Bulava:2010HSC,CLQCD:2015bgi} all predict a $J^P = 1/2^-$ and $J^P = 3/2^-$ pair of $\Omega^-$ excitations with masses lying between 1.7 and 2.2 GeV$/c^{2}$. Searching for these missing resonant excitations is essential.  

In 2018, the Belle Collaboration reported the observation of the $\Omega(2012)^-$~\cite{omega2012}, which is a candidate excited $\Omega^{-}$ state decaying into $\Xi^0 K^-$ and $\Xi^- K^0_S$. The calculations of the $\Omega^{-}$ mass spectrum in various models~\cite{Liu:2019wdr,Faustov:2015eba,Engel:2013ig,Chen:2009de,Pervin:2007wa,Oh:2007cr,Capstick:1986ter,Chao:1980em} show that the $\Omega(2012)^-$ resonance may be a good candidate for the first orbital excitation of the $\Omega^{-}$ baryon. There are also other interpretations for the $\Omega(2012)^-$, such as a hadronic molecule state~\cite{Hu:2022pae,Ikeno:2022jpe,Liu:2020yen,Ikeno:2020vqv,Lin:2019tex,Gutsche:2019eoh,Zeng:2020och,Lu:2020ste,Pavao:2018xub,Valderrama:2018bmv,Lin:2018nqd, Huang:2018wth}. In 2019, the Belle Collaboration reported an upper limit of 0.12 for the ratio of the branching ratios of $\Omega(2012)^- \to \Xi \pi \bar K$ to  $\Omega(2012)^- \to \Xi \bar K$~\cite{Belle:2019zco}, which supports the conventional interpretation. 
A later study from the Belle Collaboration reported an updated value of the ratio, i.e. $0.99 \pm 0.26 \pm 0.06$,
based on the same data sample, but with different event selection criteria and signal parametrization~\cite{Belle:2022mrg}, 
which favors the molecular interpretation of the $\Omega(2012)^-$.
The nature of the $\Omega(2012)^-$ is still controversial.  
Searching for the $\Omega(2012)^-$ and other new $\Omega^{-}$ excitations in different experiments, such as BESIII, will aid in understanding the nature of the $\Omega(2012)^-$ and other related states. 

In this Letter, we search for $\Omega^{*-}$ through the process $e^+ e^- \to \Omega^{*-} \bar{\Omega}^{+} +c.c.$, using $e^+e^-$ collision data corresponding to a total integrated luminosity of 19\,fb$^{-1}$, collected by the BESIII detector at center-of-mass energies ranging from 4.13 to 4.70\,GeV~\cite{dataset}.
Our findings present the first evidence for a new excited $\Omega^-$ hyperon, designated as $\Omega(2109)^{-}$, along with evidence for a new production mechanism for the previously identified $\Omega(2012)^-$.

Details about the design and performance of the BESIII detector are
given in Refs.~\cite{Ablikim:2009aa,Ablikim:2019hff}. 
Simulated data samples produced with a {\sc Geant4}-based~\cite{geant4} Monte Carlo (MC) package, which
includes the geometric description of the BESIII detector and the
detector response, are used to determine detection efficiencies
and to estimate backgrounds. The simulation models the beam
energy spread and initial-state radiation (ISR) in the $e^+e^-$
annihilations with the generator {\sc kkmc}~\cite{ref:kkmc}.
The inclusive MC sample includes the production of open charm
processes, the ISR production of vector charmonium(like) states,
and the continuum processes incorporated in {\sc kkmc}~\cite{ref:kkmc}.
All particle decays are modeled with {\sc evtgen}~\cite{ref:evtgen} using branching fractions
either taken from the PDG~\cite{pdg2024}, when available,
or otherwise estimated with {\sc lundcharm}~\cite{ref:lundcharm}.
Final state radiation from charged final state particles is incorporated using the {\sc photos} package~\cite{photos2}.


We search for potential excited $\Omega^-$ states in the recoil mass spectrum of the $\bar{\Omega}^+$ in the $e^+ e^- \to \Omega^{*-} \bar{\Omega}^{+}$ process, with $\bar{\Omega}^+ \to \bar{\Lambda} K^+$ and $\bar{\Lambda} \to \bar{p} \pi^+$. 
Unless explicitly stated, charge-conjugate modes are always implied throughout this Letter.
Charged tracks detected in the multilayer drift chamber (MDC) are required to be within a polar angle ($\theta$) range of $|\rm{cos\theta}|<0.93$, where $\theta$ is defined with respect to the $z$ axis,
which is the symmetry axis of the MDC. 
Particle identification for charged tracks combines measurements of the energy deposited in the MDC and the flight time in the time-of-flight
system to form likelihoods $\mathcal{L}(h)~(h=p,K,\pi)$ for each hadron $h$ hypothesis.
Each track is assigned to the particle type with the highest $\mathcal{L}(h)$. 
Events with at least one $\bar{p}$, one $K^+$, and one $\pi^+$ are kept for further analysis.

\begin{figure}[htp]
  \centering
\includegraphics[width=1\linewidth]{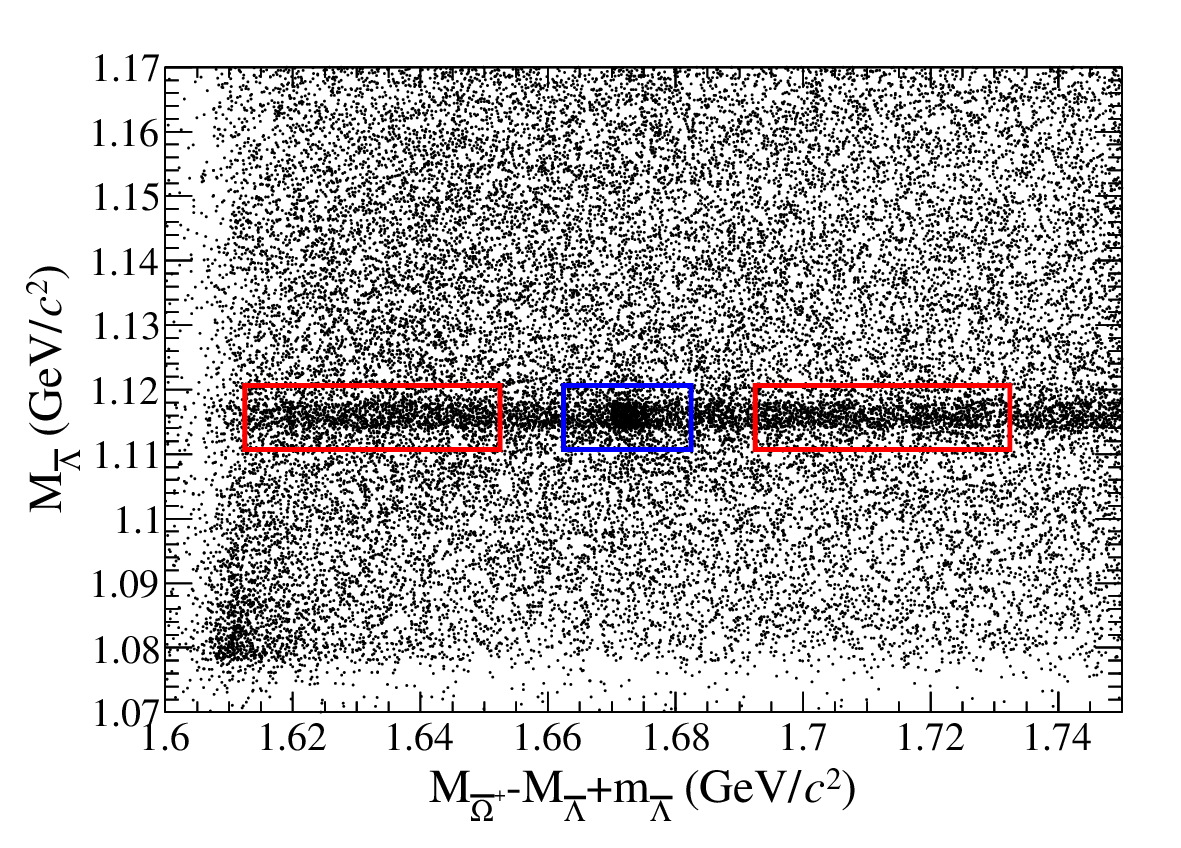}
  \caption{\small
  Distribution of $M_{\bar{\Lambda}}$ versus $M_{\bar{\Omega}^{+}}-M_{\bar{\Lambda}}+m_{\bar{\Lambda}}$ for data.
  The blue box represents the signal region, and the red boxes indicate the sideband regions.
  }
\label{fig_2d_lmd_omega}
\end{figure}

The $\bar{\Lambda}$ ($\bar{\Omega}^+$) candidates are reconstructed from $\bar{p}\pi^+$ ($\bar{\Lambda}K^+$)
pairs that are constrained to originate from a common vertex by requiring the $\chi^2$ of the vertex fits to be less
than 50. The reconstructed $\bar{\Lambda}$ invariant mass, $M_{\bar{\Lambda}}$, is required to be within $0.005\,\textrm{GeV}/c^{2}$ of the nominal $m_{\bar{\Lambda}}$ mass~\cite{pdg2024}, and the $\bar{\Omega}^+$ invariant mass, calculated using $M_{\bar{\Omega}^{+}}-M_{\bar{\Lambda}}+m_{\bar{\Lambda}}$, is required to be within $0.010\,\textrm{GeV}/c^{2}$ of the nominal $m_{\bar{\Omega}}$ mass~\cite{pdg2024}, where $M_{\bar{\Omega}^{+}}$ is the reconstructed $\bar{\Omega}^+$ mass. 
The distance between the decay vertices of the $\bar{\Lambda}$ ($\bar{\Omega}^+$) candidates and the $e^+e^-$ interaction point is required to be greater than twice the fitted uncertainties.
If there are multiple combinations, all $\bar{\Lambda}$ candidates are kept, while among the $\bar{\Omega}^+$ candidates, the one with the minimum vertex fit $\chi^2$ is retained for further analysis.
The 2D-distribution of reconstructed mass of the $\bar{\Lambda}$ candidate, $M_{\bar{\Lambda}}$,  vs.~the invariant mass of the $\bar{\Omega}^+$ candidate, $M_{\bar{\Omega}^{+}}-M_{\bar{\Lambda}}+m_{\bar{\Lambda}}$, is shown in Fig.~\ref{fig_2d_lmd_omega} for data summed over all energy points.  Clear signals of  the $\bar{\Lambda}$ and the $\bar{\Omega}^+$ can be seen. The sideband regions are chosen as $|M_{\bar{\Omega}^{+}}-M_{\bar{\Lambda}}+m_{\bar{\Lambda}}-m_{\bar{\Omega}^{+}}| \in [0.02, 0.06]\,{\rm GeV}/c^{2}$ and $|M_{\bar{\Lambda}}-m_{\bar{\Lambda}}|< 0.005\,{\rm GeV}/c^{2}$.  

\begin{figure}[htp]
  \centering
  \subfigure{\includegraphics[width=1.0\linewidth]{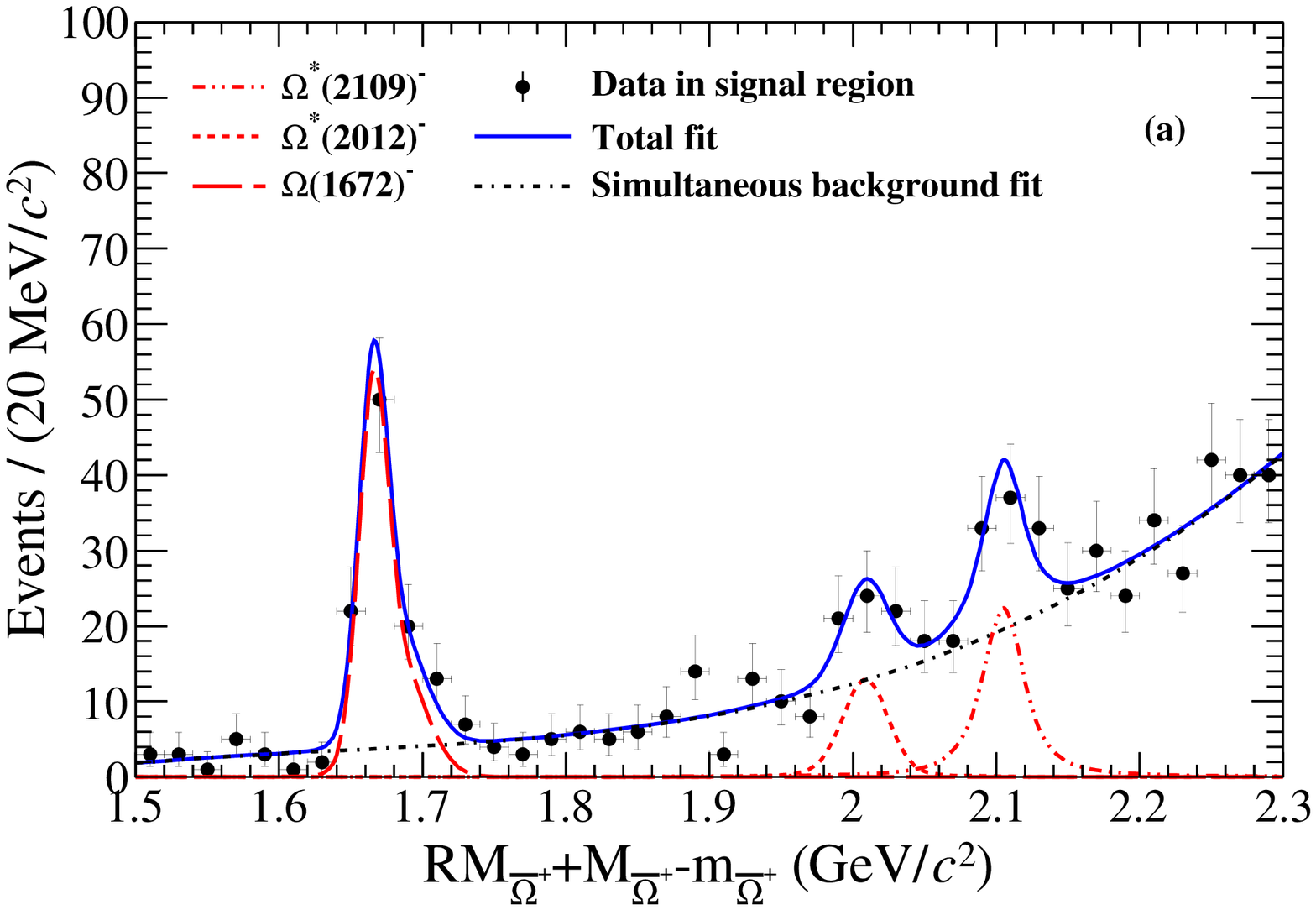}
  \label{fig_recoil_mass_signal_region}
  }
  \subfigure{\includegraphics[width=1.0\linewidth]{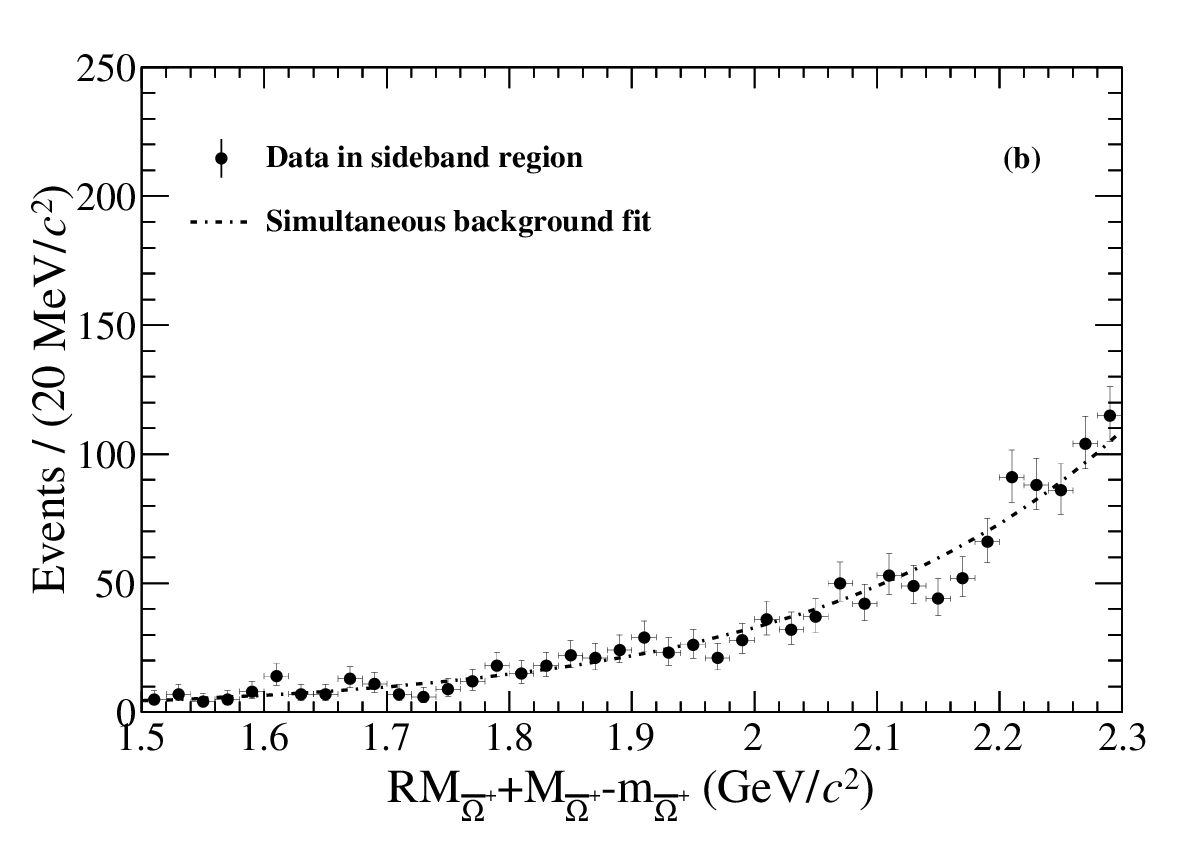}
  \label{fig_recoil_mass_sideband_region}
  }   
  \caption{\small
  Simultaneous unbinned maximum likelihood fit  the $RM_{\bar{\Omega}^+}+M_{\bar{\Omega}^+}-m_{\bar{\Omega}^+}$ distributions 
  from the signal (a) and sideband (b) regions.
  The black dots with error bars represent data. The blue solid curve is the total fit result. The red long dashed, red short dashed, and red double-dot-dashed lines are
  the signal shapes for $\Omega^{-}$, $\Omega(2012)^{-}$, and $\Omega(2109)^{-}$, respectively. The black dot-dashed line is the background shape.
  }
\label{fig_recoil_mass}
\end{figure}

Figure~\ref{fig_recoil_mass_signal_region} shows the $\bar{\Omega}^+$ recoil-mass spectrum for the candidate events falling into the signal region in Fig.~\ref{fig_2d_lmd_omega}, calculated using $RM_{\bar{\Omega}^+}+M_{\bar{\Omega}^+}-m_{\bar{\Omega}^+}$.
Here, $RM_{\bar{\Omega}^+}=|p_{e^+e^-}-p_{\bar{\Omega}^+}|$, where $p_{e^+e^-}$ is the four-momentum of the initial
$e^+e^-$ system and $p_{\bar{\Omega}^+}$ is the four-momentum of the $\bar{\Omega}^+$. The variable $RM_{\bar{\Omega}^+}+M_{\bar{\Omega}^+}-m_{\bar{\Omega}^+}$ 
provides better resolution compared to $RM_{\bar{\Omega}^+}$~\cite{zcs}.  
A clear $\Omega^{-}$ ground state signal near $1.67\,{\rm GeV}/c^{2}$ is seen, plus two additional peaks near $2.0\,{\rm GeV}/c^{2}$ and $2.1\,{\rm GeV}/c^{2}$, which we refer to as the $\Omega(2012)^{-}$ and $\Omega(2109)^{-}$, respectively.
Overall, except for these three peaks, the data sideband shape describes the data well.
Since the background is almost all from combinatorial sources and the background 
 distribution is smooth in MC studies, the data sideband events are used to constrain the shape of the background in the subsequent fit.  

A simultaneous unbinned maximum likelihood fit is performed to the $\bar{\Omega}^+$ recoil-mass spectra for the candidate events from the signal [Fig.~\ref{fig_recoil_mass_signal_region}] and sideband [Fig.~\ref{fig_recoil_mass_sideband_region}] regions. 
In this fit, the signal shapes for the $\Omega(2012)^{-}$ and $\Omega(2109)^{-}$ are nonrelativistic Breit-Wigner functions convolved 
with a double-Gaussian resolution function; the $\Omega^{-}$ is described by an asymmetric double-Gaussian function, and the background
is parametrized by an exponential function.
The parameters of the Breit-Wigner function for $\Omega(2012)^{-}$ are fixed to the PDG values~\cite{pdg2024},
 the parameters of the double-Gaussian resolution functions are fixed 
according to the MC simulation, and all other parameters for the signal shapes are free. 
According to the MC simulation, the mass resolutions for the 
$\Omega^-$, $\Omega(2012)^-$, and $\Omega(2109)^-$ are about 13, 10, and 9 ${\rm MeV}/c^{2}$, respectively.
The parameter of the exponential function is free and shared for both data samples from the signal and sideband regions.

The statistical significances of the two $\Omega^{-}$ excitations are calculated by excluding the
corresponding peak from the fit, finding the change in the log-likelihood $\Delta({\rm 2ln}\mathcal{L})$,
and converting this to a $p$ value, taking into account the change in degrees of freedom ($N_{\rm{dof}}$)~\cite{Wilks:1938dza}.
This is then converted to an effective number of standard deviations $n_{\sigma}$.
For the peak around $2.0\,{\rm GeV}/c^{2}$, we obtain $\Delta({\rm 2ln}\mathcal{L}) = 12.2$ with $N_{\rm{dof}} = 1$, yielding $n_{\sigma}=3.5$.
This provides evidence for $\Omega(2012)^{-}$ production, with a signal yield of $24 \pm 8$. 
For the peak around $2.1\,{\rm GeV}/c^{2}$, we obtain $\Delta({\rm 2ln}\mathcal{L}) = 23.8$ with $N_{\rm{dof}} = 3$, yielding $n_{\sigma}=4.2$.
This gives evidence for a new excited $\Omega^{-}$ hyperon, the $\Omega(2109)^{-}$, with signal yield of $53 \pm 19$.
The fitted mass and width of the $\Omega(2109)^{-}$ are 
$m_{\Omega(2109)^{-}}=2108.5 \pm 5.2_{\rm stat}\,{\rm MeV}/c^{2}$
and
${\Gamma}_{\Omega(2109)^{-}}=18.3 \pm 16.4_{\rm stat}\,{\rm MeV}.$

Systematic uncertainties on the measurements of the mass and width for the $\Omega(2109)^{-}$ mainly originate from the background shape, the mass shift, the resolution, 
the efficiency curve, the choice of sideband, and the parametrization of the $\Omega(2012)^{-}$ and $\Omega(2109)^{-}$ signal shape. The numerical results are summarized in Table~\ref{tab_sys}.
The impact of the background shape is evaluated by varying it to be a third-order Chebyshev polynomial. The uncertainties from the mass shift and resolution are determined by calibrating the mass shift and resolution difference between data and MC simulation using the control sample of $e^+ e^- \to \Omega^{-} \bar{\Omega}^{+}$. The effects of the dependence of efficiency on the fitted variable are assigned according to the differences between the results obtained with and without consideration of the efficiency dependence. The systematic uncertainty related to the choice of sideband is estimated by ten different variations in the edges of the sideband intervals, by amounts ranging from 5 to 20\,MeV/$c^2$, and the maximum change of the mass or width compared to the nominal one is taken as the uncertainty. 
To consider the uncertainty from the parametrization of the $\Omega(2012)^{-}$ signal shape,
we change the fixed parameters of $\Omega(2012)^{-}$ adopted from PDG~\cite{pdg2024} by $\pm 1\sigma$, 
and the largest variation of the mass or width is considered as one contribution to this uncertainty;
meanwhile, the variation of the mass and width when the $\Omega(2012)^{-}$ is alternatively fitted by the signal MC shape taking into account the shape distortion of $\Omega(2012)^{-}$ due to the finite phase space in $\Omega(2012)^{-}$ three-body decays (such as $\Omega(2012)^{-} \to \Xi(1530) K \to \Xi \pi K$) is considered as another contribution to this uncertainty.
The effect from the parametrization of $\Omega(2109)^{-}$ signal shape is studied
by varying the double-Gaussian function convoluted with the Breit-Wigner to be a single-Gaussian function.
The potential contributions from the feed-down components ($e^+e^- \to \Omega^- \bar{\Omega}(2012/2109)^+,  \bar{\Omega}(2012/2109)^+ \to \bar{\Omega}^+ \pi \pi$)
and the interference between $\Omega(2012)^{-}$ and $\Omega(2109)^{-}$ are investigated, 
while no evidence of these contributions is found at the current statistical level of the data.
The changes in mass
and width when we introduce interference between $\Omega(2012)^{-}$ and $\Omega(2109)^{-}$ in the fit
are quoted as systematic uncertainties. 
Following Refs.~\cite{LHCb:2017uwr, LHCb:2023sxp}, the upper
limit on the natural width of $\Omega(2109)^-$ is set at Bayesian 95\% confidence level, assuming Gaussian behavior for both statistical and
systematic uncertainties, yielding ${\Gamma}_{\Omega(2109)^{-}}<46.9\,{\rm MeV}.$

\begin{table}[htp]
\setlength{\tabcolsep}{1mm}
\centering
\caption{Absolute systematic uncertainties for the measurements of the resonance parameters of $\Omega(2109)^{-}$.
}
\renewcommand{\arraystretch}{1.0}
\begin{tabular}{ccc}
\specialrule{0em}{1pt}{1pt}
  \hline
  \hline
  \multirow{2}{*}{Source}&						Mass&		Width\\
  &								(${\rm MeV}/c^{2}$)&	($\rm MeV$)\\
  \hline
  Background shape&				0.1&						0.1\\
  Mass shift&						0.3&						···\\
  Resolution&						···&						4.7\\
  Efficiency curve&					0.1&						0.6\\
  Choice of sideband&				0.1&						0.2\\
  Parametrization of $\Omega(2012)^{-}$&		\multirow{2}{*}{0.1}&			\multirow{2}{*}{0.3}\\
    signal shape&	&	\\
  Parametrization of $\Omega(2109)^{-}$&		\multirow{2}{*}{0.5}&			\multirow{2}{*}{2.4}\\
    signal shape&	&	\\
  Interference&						0.6&						2.0\\
\hline
Total&							0.9&						5.7\\
  \hline
  \hline
\end{tabular}
\label{tab_sys}
\end{table}

Following Ref.~\cite{zcs}, 
we take the look-elsewhere effect and the relevant systematic uncertainties into account simultaneously to reestimate the significance of the $\Omega(2109)^{-}$ using 5000 pseudodatasets to obtain the distribution of $\Delta({\rm 2ln}\mathcal{L})$.
The resulting distribution is found to be well described by a $\chi^2$ distribution with $3.27 \pm 0.03$ degrees of freedom. 
Combining with the minimum $\Delta({\rm 2ln}\mathcal{L})=23.6$ among all the systematic tests, the significance of the $\Omega(2109)^{-}$ is determined to be $n_{\sigma}=4.1$ conservatively with $N_{\rm{dof}} = 3.27 + 0.03 = 3.30$.
For the significance of the $\Omega(2012)^{-}$, to account for the systematic effects, the minimum $\Delta({\rm 2ln}\mathcal{L})$ among all systematic variations is taken, which is still 12.2. Combining with $N_{\rm{dof}} = 1$, we obtain $n_{\sigma}=3.5$ for $\Omega(2012)^{-}$ including systematic uncertainties.

Since both of the significances of $\Omega(2012)^{-}$ and $\Omega(2109)^{-}$ are in the range of [3, 5]\,$\sigma$, we report the ratios of the average cross sections of 
$e^+ e^- \to \Omega(X)^{-} \bar{\Omega}^{+}+c.c.$ to $e^+ e^- \to \Omega^{-} \bar{\Omega}^{+}$, $R_{\Omega(X)^{-}}$, 
in the range of $\sqrt{s} \in [4.13, 4.70]\,\textrm{GeV}$ and also the corresponding upper limits at the 90\% confidence level, $R_{\Omega(X)^{-}}^{\rm UL}$.
Here, $R_{\Omega(X)^{-}} \equiv \frac{\sigma(e^+ e^- \to \Omega(X)^{-} \bar{\Omega}^{+}+c.c.)}{\sigma(e^+ e^- \to \Omega^{-} \bar{\Omega}^{+})}$, with $\Omega(X)^{-}$ denoting $\Omega(2012)^{-}$ or $\Omega(2109)^{-}$, and are determined as 
\begin{equation}
R_{\Omega(X)^{-}}  \,=\,  \frac{N_{\Omega(X)^{-}}}{N_{\Omega^{-}}} 
  \left( \frac{(1+\delta^r)_{\Omega^{-}}}{(1+\delta^r)_{\Omega(X)^{-}}} \right)
  \left( \frac{\epsilon_{\Omega^{-}}}{\epsilon_{\Omega(X)^{-}}} \right) ,
\label{eq_cs_2109_all}
\end{equation}
where $N_{\Omega(X)^{-}}$ and $N_{\Omega^{-}}$ represent the numbers of events, 
$(1+\delta^r)_{\Omega(X)^{-}}$ and $(1+\delta^r)_{\Omega^{-}}$ represent the ISR correction factors, and 
$\epsilon_{\Omega(X)^{-}}$ and $\epsilon_{\Omega^{-}}$ represent the detection efficiencies 
 for $e^+ e^- \to \Omega(X)^{-} \bar{\Omega}^{+}$ and $e^+ e^- \to \Omega^{-} \bar{\Omega}^{+}$, respectively.
$\frac{N_{\Omega(X)^{-}}}{N_{\Omega^{-}}}$ are determined to be $0.26\pm0.09$ and $0.58\pm0.22$ by the fit shown in Fig.~\ref{fig_recoil_mass}, 
$\frac{(1+\delta^r)_{\Omega^{-}}}{(1+\delta^r)_{\Omega(X)^{-}}}$ are calculated to be 1.11 and 1.17 by {\sc kkmc}~\cite{ref:kkmc} with
the cross-section line shape taken from Ref~\cite{bam586}, 
and $\frac{\epsilon_{\Omega^{-}}}{\epsilon_{\Omega(X)^{-}}}$ are evaluated to be  $1.23 \pm 0.01$ and $1.38 \pm 0.01$ with phase space MC simulation for the $\Omega(2012)^{-}$ and $\Omega(2109)^{-}$, respectively.
Based on these values and Eq.~(\ref{eq_cs_2109_all}), we obtain $R_{\Omega(2012)^{-}}=0.36\pm0.13_{\rm stat}$ and $R_{\Omega(2109)^{-}}=0.93\pm0.35_{\rm stat}$.

The systematic uncertainties for $\frac{N_{\Omega(X)^{-}}}{N_{\Omega^{-}}}$ come from sources similar to those for the measurements of the $\Omega(2109)^{-}$ parameters, including the background shape, the resolution, the efficiency curve, the choice of sideband, and the parametrization of the $\Omega(2012)^{-}$ signal shape. Thus, the same variation methods are adopted to estimate the corresponding effects here.  
Following the same method used in Ref.~\cite{BESIII:2024ext}, the systematic uncertainties for $\frac{(1+\delta^r)_{\Omega^{-}}}{(1+\delta^r)_{\Omega(X)^{-}}}$ are assigned by changing the input cross-section line shape to the phase-space line shape. 
The systematic uncertainties associated with $\frac{\epsilon_{\Omega^{-}}}{\epsilon_{\Omega(X)^{-}}}$ are examined from two perspectives:
(1) the differences in $\frac{\epsilon_{\Omega^{-}}}{\epsilon_{\Omega(X)^{-}}}$ obtained from two different MC samples, with the recoil $\Omega(X)^{-}$, $\Omega^-$ decaying invisibly or exclusively to $\Xi K/\Lambda K$; 
(2) the variations in $\frac{\epsilon_{\Omega^{-}}}{\epsilon_{\Omega(X)^{-}}}$ by changing the $\bar{\Omega}^{+}$ production angular model in the MC simulation to 
align with the expected distribution in a generalized-vector-meson dominance model~\cite{Korner:1986vi}, 
assuming the $J^P$ for $\Omega(2102)^-$ and $\Omega(2109)^-$ are $1/2^-$ and $3/2^-$, respectively. 
The corresponding numerical values are summarized in Table~\ref{tab_sys_R}.

\begin{table}[htp]
\setlength{\tabcolsep}{6mm}
\centering
\caption{Systematic uncertainties in \% for $R_{\Omega(X)^{-}}$.
}
\renewcommand{\arraystretch}{1.0}
\begin{tabular}{ccc}
\specialrule{0em}{1pt}{1pt}
  \hline
  \hline
  Source&													$R_{\Omega(2012)^{-}}$&		$R_{\Omega(2109)^{-}}$\\
  \hline
  $\frac{N_{\Omega(X)^{-}}}{N_{\Omega^{-}}}$&					1.8&				3.1\\
  \rule{0pt}{15pt}
  $\frac{(1+\delta^r)_{\Omega^{-}}}{(1+\delta^r)_{\Omega(X)^{-}}}$&	7.9&				11.4\\
  \rule{0pt}{15pt}
  $\frac{\epsilon_{\Omega^{-}}}{\epsilon_{\Omega(X)^{-}}}$&			5.9&				6.5\\
\hline
Total&													10.0&			13.5\\
  \hline
  \hline
\end{tabular}
\label{tab_sys_R}
\end{table}

Based on a Bayesian method~\cite{bayesian},  the raw likelihood distributions versus the corresponding $R_{\Omega(X)^{-}}$ are obtained.
Then, the likelihood distribution after considering the additive systematic uncertainty is smeared with a Gaussian function with the mean of zero and the width of the total multiplicative systematic uncertainty. Based on this, $R_{\Omega(2012)^{-}}^{\rm UL}$ and $R_{\Omega(2109)^{-}}^{\rm UL}$ are set to be 0.57 and 1.31, respectively.


In summary, using data samples with an integrated luminosity of 19\,fb$^{-1}$ collected by the BESIII detector at center-of-mass energies ranging from 4.13 to 4.70\,GeV,
we find an evidence for the $\Omega(2109)^{-}$ hyperon through the process $e^+ e^- \to \Omega(2109)^{-} \bar{\Omega}^{+}$ with a significance of 4.1\,$\sigma$ including both the look-elsewhere effect and systematic uncertainties.
The mass and width of $\Omega(2109)^{-}$ are measured to be
$2108.5 \pm 5.2_{\rm stat} \pm 0.9_{\rm syst}\,{\rm MeV}/c^{2}$
and
$18.3 \pm 16.4_{\rm stat} \pm 5.7_{\rm syst}\,{\rm MeV}$, respectively.
The upper limit on the natural width of $\Omega(2109)^-$ is set to be $46.9\,{\rm MeV}$ at 95\% confidence level.
We also find evidence for $\Omega(2012)^{-}$ production via the process $e^+ e^- \to \Omega(2012)^{-} \bar{\Omega}^{+}$, 
with a significance of 3.5\,$\sigma$ including systematic uncertainties. 
Furthermore, the ratios $R_{\Omega(X)^{-}}$ of the average cross sections of 
$e^+ e^- \to \Omega(X)^{-} \bar{\Omega}^{+}+c.c.$ and $e^+ e^- \to \Omega^{-} \bar{\Omega}^{+}$ in the range of $\sqrt{s} \in [4.13, 4.70]\,\textrm{GeV}$ are measured as 
$R_{\Omega(2012)^{-}}=0.36\pm0.13_{\rm stat}\pm0.04_{\rm syst}$ and $R_{\Omega(2109)^{-}}=0.93\pm0.35_{\rm stat}\pm0.13_{\rm syst}$,
while the corresponding upper limits at the 90\% confidence level are set at $R_{\Omega(2012)^{-}}^{\rm UL}=0.57$ and $R_{\Omega(2109)^{-}}^{\rm UL}=1.31$.

Evidence for the $\Omega(2109)^{-}$ and the $\Omega(2012)^{-}$ reported in this Letter helps to fill in the missing excited states in the $\Omega^-$ hyperon spectrum. 
Theoretical models~\cite{Oh:2007cr, Capstick:1986ter, Faustov:2015eba, Chao:1980em, Chen:2009de, Pervin:2007wa, Liu:2019wdr, An:2013zoa, An:2014lga, Engel:2013ig, Edwards:2012fx, Bulava:2010HSC,CLQCD:2015bgi} predict a $J^P = 1/2^-$ and $J^P = 3/2^-$ pair of $\Omega^-$ excitations within the mass region from 1.7 to 2.2 GeV$/c^{2}$, but with large variations in their mass predictions.
We note that our results are very close to the lattice QCD calculation reported by the Hadron Spectrum Collaboration~\cite{Edwards:2012fx} where the masses of the $J^P = 1/2^-$ and $J^P = 3/2^-$ pair of $\Omega^-$ excitations are predicted to be around 2.0 GeV$/c^{2}$ and 2.1 GeV$/c^{2}$, respectively.
This consistency would imply that the $\Omega(2109)^{-}$ and $\Omega(2012)^{-}$ are probably conventional three-quark baryons and not molecules or other exotic states. 
The BEPCII upgrade, already implemented in 2024, will lead to much higher luminosity, allowing  the confirmation of and parameter measurements for these two states to be investigated with larger data samples.
\acknowledgments
The BESIII Collaboration thanks the staff of BEPCII and the IHEP computing center for their strong support. This work is supported in part by National Key R\&D Program of China under Contracts Nos. 2023YFA1606704, 2023YFA1609400, 2020YFA0406300, 2020YFA0406400, 2023YFA1606000; National Natural Science Foundation of China (NSFC) under Contracts Nos. 12375092, 12105127, 11635010, 11735014, 11935015, 11935016, 11935018, 12025502, 12035009, 12035013, 12061131003, 12192260, 12192261, 12192262, 12192263, 12192264, 12192265, 12221005, 12225509, 12235017, 12361141819; the Chinese Academy of Sciences (CAS) Large-Scale Scientific Facility Program; the CAS Center for Excellence in Particle Physics (CCEPP); Joint Large-Scale Scientific Facility Funds of the NSFC and CAS under Contract No. U1832207; 100 Talents Program of CAS; The Institute of Nuclear and Particle Physics (INPAC) and Shanghai Key Laboratory for Particle Physics and Cosmology; German Research Foundation DFG under Contracts Nos. FOR5327, GRK 2149; Istituto Nazionale di Fisica Nucleare, Italy; Knut and Alice Wallenberg Foundation under Contracts Nos. 2021.0174, 2021.0299; Ministry of Development of Turkey under Contract No. DPT2006K-120470; National Research Foundation of Korea under Contract No. NRF-2022R1A2C1092335; National Science and Technology fund of Mongolia; National Science Research and Innovation Fund (NSRF) via the Program Management Unit for Human Resources \& Institutional Development, Research and Innovation of Thailand under Contracts Nos. B16F640076, B50G670107; Polish National Science Centre under Contract No. 2019/35/O/ST2/02907; Swedish Research Council under Contract No. 2019.04595; The Swedish Foundation for International Cooperation in Research and Higher Education under Contract No. CH2018-7756; U. S. Department of Energy under Contract No. DE-FG02-05ER41374.


\end{document}